\documentclass[preprint,aps]{revtex4}
\begin{document}
\title{Lattice model of gas condensation within
nanopores}
\author{Raluca A. Trasca, M. Mercedes Calbi and Milton W.Cole}
\address{Department of Physics,Pennsylvania State University,University Park,
PA 16802}

\begin{abstract}
We explore the thermodynamic behavior of gases
adsorbed within a nanopore. The theoretical
description employs a simple lattice gas model, with
two species of site, expected to describe various
regimes of  adsorption and condensation behavior. The
model includes four hypothetical phases: a cylindrical shell phase $(S)$, in
which the sites close to the cylindrical wall are
occupied, an
axial phase $(A)$, in which sites along the cylinder's
axis are occupied, a full phase $(F)$, in which all sites are
occupied, and an empty phase $(E)$. We obtain exact results
at $T=0$ for the phase behavior, which is a function of
the interactions present in any specific problem.
We obtain the corresponding results at finite $T$
from mean field theory. Finally, we examine the model's predicted phase
behavior of some real gases adsorbed in nanopores.
\end{abstract}

\maketitle

\section{INTRODUCTION}

\label{sec:introduc}

A focus of current attention in statistical physics is
the behavior of matter in confining geometries
\cite{bar,guyer,tal,page,sar,liu,chan,thom,gubb,schoen,art,evans}. An
extreme version of this problem arises for adsorption
within nanotubes, a case for which the transverse
dimensions may be of order molecular sizes. One
expects an important parameter in this class of
problem to be the ratio $R^*$ of the diameter of the
molecule to that of the tube. When this ratio is of
the order one, the
adsorbate may be well described by a one-dimensional
(1D) model. As $R^*$ decreases, one expects there to
arise successively a sequence of onion-like concentric
shells of matter; the number of possible shells is
critically dependent on the value of $R^*$. Accompanying
the variation in $R^*$ is a variation of energy scales,
which are the crucial variables in the thermodynamics
of the system.

There have been performed many studies of specific
geometries and specific adsorbate-substrate
combinations, as recently reviewed by Gelb et al
\cite{gelb}. However, there have been relatively few
studies undertaken of the general problem of
adsorption in pores in the case of variable $R^*$. The
present work represents an effort in that direction.
Here, we employ a highly oversimplified lattice model
of adsorption \cite{ebner,olive} designed for cases when one or two
concentric phases of matter (but no more) may be
present. Since the present analysis is limited by the
assumption of just two distinct species of lattice
sites, it describes just the $R^* \geq 1$ regime. Hence,
there are assumed to be four possible phases for this
geometry: an empty phase $(E)$, an axial phase $(A)$,
in which atoms are adsorbed only on the cylinder's axis,
a cylindrical shell phase $(S)$, in which atoms condense
close to the cylinder's wall, and a full phase $(F)$, in which
both axial and shell sites are populated with atoms. These are 
depicted schematically in Fig.1. We assume a
model which includes both pore-site interactions and nearest
neighbor interactions. Since the pore attraction is
usually different for shell and axial sites, we may think of the axial
and shell atoms as two different species interacting each other with a 
common value of the chemical potential $\mu$. The same idea was explored
in adsorption problems involving two types of binding sites \cite{machta}. 
Besides the pore attraction, the atoms experience an intra-species
interaction (axial-axial or shell-shell) and an inter-species
interaction (axial-shell). The phase behavior depends on the
values of these various energies, especially on the
attractive or repulsive character of the inter-species
interaction.

Section $2$ of this paper presents results at zero temperature $(T)$
for the exact
phase behavior as a function of the interactions. Section $3$
reports a mean field evaluation of the phase behavior
at finite $T$. The adsorption behavior given by finite $T$
isotherms is compared to the phase diagrams at $T=0$.
Section $4$ describes the relation between the lattice models and
some examples of possible realistic situations, i.e., gases
adsorbed in carbon nanotubes of various radii.
Ultimately, we would like to relate the systems' properties to energy
scales present in the real problem. Since these may
not be known, it becomes possible in principle to deduce these by
comparing experimentally observed phase behavior with
that predicted by the model.
In view of the approximations inherent in the lattice model,
we believe that our results provide a qualitative
picture of the expected phase behavior and its
evolution with the size ratio $R^*$ mentioned above.

\section{ZERO TEMPERATURE ANALYSIS}

	As a starting point, we consider adsorption in infinite cylindrical pores 
at $T=0$. The possible phases are described in the introduction. The cases when the shell-axial interaction is attractive or
repulsive are investigated separately. We will illustrate in detail our
analysis for the case of an attractive inter-species interaction. Initially, to simplify
the discussion, assume that the
analysis can be divided into two alternative approaches. In
one, we consider the only possible phases to be $E,A$ and $F$. In the other, we
consider just the phases $E,S$ and $F$. We show below that this separation into two
distinct treatments encompasses all possibilities for the case of attractive interactions
between $A$ and $S$ sites. However, in the case of a repulsive interaction, this
division of the problem into two parts does not work, necessitating a somewhat more
complicated numerical analysis.

 	The phase transition diagrams are constructed on the basis of free
energy considerations. The shell species is
adsorbed on a 2D lattice of sites, with the interaction energy
$\epsilon_s$ between particles at adjacent sites. For simplicity, this 2D
lattice on a cylindrical surface is taken as a square lattice; hence, the number of nearest shell
neighbors of a shell atom $z_s$ is $4$. The axial species is adsorbed on a
1D lattice of sites, of interaction strength $\epsilon_a$ and
coordination number $z_a = 2$. We include also the interaction between axial
and shell sites, denoted by $\epsilon_{sa}$. Throughout
the paper, we express all energies, chemical potentials and temperatures
in units of $\epsilon = \epsilon_s = \epsilon_a$ , the absolute value of the
interatomic interaction. For simplicity, we assume that shell atoms sit on rings whose centers are
occupied by axial atoms. The number of axial neighbors for a
shell site $(z_{sa})$ is $1$ and the number of shell neighbors for an
axial site $(z_{as})$ is larger than $1$.

 	We first determine the equilibrium phase as a function of $\mu$. The axial,
shell and full grand free energies ($\Omega = F-\mu N$, where $F$ is the Helmholtz
free energy) at $T = 0$ , can be written as:
\begin{equation} 
\Omega_{a} = N_a ( V_a - \frac{z_a}{2}) - \mu N_a
\end{equation}	                            
\begin{equation} 
\Omega_s= N_s(V_s - \frac{z_s}{2})- \mu N_s
\end{equation}
\begin{equation}
\Omega_f= N_a(V_a - \frac{z_a}{2}) + N_s(V_s - \frac{z_s}{2})+ N_s z_{sa}\epsilon_{sa}- \mu(N_a + N_s)
\end{equation}
where $N_{a(s)}$ is the number of sites in the axial (shell) phase and $V_{a(s)}$ is the
interaction potential energy experienced by the axial (shell) site due to the nanotube environment.
Adsorption in nanopores at $T=0$ can occur only if the
adsorbate is attracted to the interior of the nanopore, i.e.
$V_{a(s)}<0$. We denote the ratio of axial to
shell densities (number of atoms per pore length) as $\gamma =N_a/N_s$. The
axial and shell cohesive energies per particle are respectively:
\begin{equation}
E_a=-(V_a-z_a/2)
\end{equation}
\begin{equation}
E_s=-(V_s-z_s/2)
\end{equation}
These energies consist of the pore attraction energy and the nearest neighbor
interaction (the factor of $1/2$ avoids double counting). With this notation, and
replacing $z_{sa}$ by $1$, the grand free energies can be rewritten as:
\begin{equation}
 \frac{\Omega_a}{N_s}=-\gamma E_a-\gamma \mu
\end{equation}
\begin{equation}
 \frac{\Omega_s}{N_s}=-E_s-\mu
\end{equation} 
\begin{equation}                              
\frac{\Omega_f}{N_s}=-\gamma E_a - E_s  +\epsilon_{sa} - \mu(\gamma+1)
\end{equation}             
	One observes that the adsorption behavior (as a function of $\mu$) depends on four 
parameters: $\gamma$, $\epsilon_{sa}$, $E_a$ and $E_s$. The $T=0$ isotherms are determined 
by finding the minimum of these $\Omega$ values and comparing the result with the empty lattice 
result $\Omega_E=0$.
  	 The axial phase is favored relative to the empty phase if $\Omega_a < 0$, i.e.
\begin{equation}
\mu > -E_a 
\end{equation}                                                               
 	The full phase is lower in grand free energy than the empty phase if $\Omega_f < 0$, i.e.
\begin{equation}
\mu>(-E_s - E_a \gamma+\epsilon_{sa})/(1+\gamma)
\end{equation}
	The  axial phase is favored relative to the full phase if
$\Omega_a < \Omega_f$, implying
\begin{equation}
\mu < -E_s + \epsilon_{sa}
\end{equation}
  	An analogous argument is true for the shell phase. $\Omega_s<0$ implies	
\begin{equation}
\mu > -E_s
\end{equation}
Note that $\Omega_s < \Omega_f$ if
\begin{equation}                                     
\mu<-E_a + \frac{\epsilon_{sa}}{\gamma}
\end{equation}
                                                                          
	First, we construct two independent phase diagrams with $E_{a(s)}$ and
$\mu$ as coordinates, corresponding to $(E,S,F)$ and $(E,A,F)$ cases.
Then, by inspecting the diagrams, we learn how to combine them into a 
single diagram applicable to both
cases at once. 	We first analyze the $E, S , F$ possible phase
transitions alone.
The $\mu$ regime of each phase is determined by comparison using the equations $(10),(12),(13)$.
The transitions between these phases occur at values of 
$\mu$ such that the inequalities $(10),(12),13)$ become equalities.
In addition, we have to take into
consideration that the chemical potential of the pore condensation should be
smaller than the chemical potential of bulk condensation in the simple cubic
lattice Ising model, which is $\mu_0 = -3$. (Of course, transitions can occur 
within the pore for $\mu > \mu_0$, but one does not ordinarily study them.) 
Due to this restriction,
we can distinguish two cases. The first occurs when the $S \leftrightarrow F$ transition
is below saturation ($-E_a+\epsilon_{sa}/\gamma<-3$).
Then, all three phases $E, S, F$ are possible, as shown in $Fig.2a$. The alternative 
scenario occurs when
the $S \leftrightarrow F$ transition is above saturation
($-E_a+\epsilon_{sa}/\gamma>-3$). In this case, there are only two possible phases, $E$
and $S$, as shown in $Fig.2b$.

 	The $E , A , F$ phase analysis is very similar
to that above for $E , S , F$.  The two cases which can be distinguished
here are : $a) -E_s + \epsilon_{sa}>-3$, when all three phases
($E, A, F$) are possible and $b)-E_s+\epsilon_{sa}<-3$ , when there are only two possible
phases, $E$ and $A$.

  	So far, the phase transition behavior has been derived from two separate analyses: $E, S , F$ and $E , A, F$.
We now show how the parameter values may be assessed in order to establish which of
the two analyses is appropriate to a given system, i.e. a specified set of parameters. \
To do so, we need to compare values of
$\Omega_a$ and $\Omega_s$. 
The difference between the relevant free
energies satisfies:
 \begin{equation}
\frac{\Omega_s - \Omega_a}{N_s} = - E_s + E_a \gamma - \mu(1-\gamma)
\end{equation}
As can be seen from $eqn.9$ , the $E \leftrightarrow A$ transition occurs at
$\mu_{ea} = -E_a$. In the limit $\mu=\mu_{ea}$, then
\begin{equation}
 \frac{\Omega_s - \Omega_a} {N_s} = -(E_s - E_a)
\end{equation}                            
If $E_a < E_s $, then $\Omega_s < \Omega_a$ at this value of $\mu$. At higher value of
$\mu$ ($>\mu_{ea}$), $\Omega_s$ remains less than $\Omega_a$. Hence a transition to the 
axial phase does not
occur for any $\mu$. If, instead, $E_s<E_a$, then
$\Omega_a<\Omega_s$ and the axial phase is stable at $\mu=\mu_{ea}$. Is it possible that
$\Omega_s - \Omega_a$ changes sign for higher $\mu$ (corresponding to an $A$ to $S$
transition)? This would require $\Omega_a = \Omega_s$ at a transition value $\mu=\mu_{as}$
such that
\begin{equation}
\mu_{as}=\frac{E_a (\gamma - \rho)}{1- \gamma}
\end{equation}
where $\rho=E_a/E_s <1$. Hence $\rho-\gamma>3(1-\gamma)/E_a>1-\gamma$. This implies
$\rho>1$, which violates the assumption $E_s<E_a$. This rules out such a possibility. 

	The same examination can be done at the $E \leftrightarrow S$
transition line, $\mu_{es} = -E_s$; we then find that for $E_s < E_a$, the shell
phase does not occur. Hence the possibilities are either $E_s > E_a$ (never
the $A$ phase) or $E_a > E_s$ (never the $S$ phase). This justifies the separate
analyses used above for the two distinct cases which can arise.

Because the two cases correspond to different regimes of parameter space, $E_s > E_a$ and
$E_s < E_a$, they can be merged in a phase diagram which has as coordinates the
interactions present in our problem: $E_a$ and $E_s$. One has only to analyze $Fig.2a,b$
and find the adsorption sequences as a function of both interactions when $\mu$ is
increased. $Fig.3$ exhibits the regimes of distinct adsorption sequences. All possible
sequences occur except those ruled out by the thermodynamic stability condition
$\partial \mu /\partial N>0$. The region denoted $E$
corresponds to repulsive, or weakly attractive, pore-gas interactions, so that no atoms adsorb inside the
pore. In the $E\rightarrow A$ region, the shell phase's chemical potential of
condensation is greater than $- 3$, so the F phase does not occur. Physically, the $E\rightarrow A$
region corresponds to a repulsive, or weakly attractive, pore-shell interaction and an attractive
pore-axis interaction; hence, atoms adsorb only at the axial sites . In
the $E\rightarrow A\rightarrow F$ region, the attraction in the
axial phase
is larger than that in the shell phase, so that the axial region is
occupied first and then the shell follows at higher $\mu$. Similar reasoning applies to the $E\rightarrow S$
and $E\rightarrow S\rightarrow F$ regions. Possibly, the most interesting behavior occurs
in the $E\rightarrow F$ region. In general, as seen more clearly at finite 
$T$, the axial and shell condensations occur at
different chemical potentials. However, in
the case of an attractive axial-shell interaction, when the shell and axial
energies per particle are similar, the shell and axial phases
become cooperative and undergo a common pore filling transition.

	We have examined thus far the case of an attractive axial-shell interaction. In the
repulsive case, the inter-species interaction energy ($\epsilon_{sa}$) is positive.
Then, we have to take into account a new possibility, the
transition from axial to shell phase (alone). Physically, this means that when
the shell atoms are adsorbed, the axial phase, which has a lower density than
the shell phase, is expelled by the repulsive axial-shell interaction. Therefore,
we compare all the grand free energies $\Omega_a,\Omega_s,\Omega_f$ with
each other and the zero energy of the E phase. We present the resulting phase
diagrams in $(E_a,\mu)$ and  $(E_s,\mu)$ coordinates in $Fig.4a$ and $4b$. Both diagrams
exhibit all phases and possible transitions $E\leftrightarrow A, E\leftrightarrow S ,
 A\leftrightarrow S , A\leftrightarrow F$ and $S\leftrightarrow F$, but there is no
$E \leftrightarrow F$ transition.
There are several qualitative differences between this case, shown in $fig.5$, and the attractive interaction
case, shown in $fig.3$. Missing in the repulsive case is $E \leftrightarrow F$; present
in this case are $E \leftrightarrow A \leftrightarrow S \leftrightarrow F$ and
$E \leftrightarrow A \leftrightarrow S$ sequences (absent in the attractive case).
The last two are associated with the appearance of $S$, at the expense of A atoms,
in order to decrease $\Omega$ by adding more particles.

\section{FINITE TEMPERATURE ANALYSIS}

	In this section, we explore the phase transitions at finite $T$
for a gas within our pore. This is a 1D system in the thermodynamic limit of
divergent length. To study this model, we use mean field theory. It is known
that 1D systems do not exhibit
 phase transitions at any finite T. However, in the present mean
field treatment, we obtain a spurious transition. The results of an exact
calculation of the phase behavior in a square pore \cite{swift} were found to be $qualitatively$ similar to those of mean field
theory, apart from a narrow regime of $\mu$ where spurious transitions occur in mean
field theory; these are replaced by nearly discontinuous isotherms in the exact case. We
note that gases in some nanoporous media (zeolites or nanotube bundles) may represent
quasi-1D systems which can go
through a genuine phase transition when molecules in adjacent pores are coupled. This transition has been studied recently in a
number of models of gases in pores, by both simulations and exact models\cite{rad,cole,calbi,fisher}.
  	
  	The occupation probabilities of axial and shell sites are called $n_a$ and $n_s$, respectively.
We construct the grand free energy of the system and minimize it with
respect to $n_s$ and $n_a$.
The same procedure was used in Refs.8 and 9 for analyzing layering and wetting
phase transitions.  	
  	The energy $U$ of the system is a generalization to finite T of the calculation in
Section $2$. Specifically, the energy is:
\begin{equation}
U=N_s n_s(-\frac{z_s}{2} n_s+V_s )+N_a n_a(-\frac{z_a}{2} n_a+V_a)+ N_s n_s(z_{sa}n_a\epsilon_{sa})
\end{equation}
and the entropy is written as:
\begin{equation} 	
S=-N_s[n_s \ln n_s+(1-n_s)\ln(1-n_s)]-N_a[n_a\ln n_a+(1-n_a)\ln(1-n_a)]
\end{equation}
	
	The minimization of the grand free energy
$ U-TS-\mu N $ with respect
to the occupation numbers $n_a$ and $n_s$ yields two coupled equations, as found in
reference [8]:
\begin{eqnarray}
n_s &=&\frac {1}{1+\exp(-\beta(\mu-V_s-z_s\epsilon_s n_s-z_{sa}\epsilon_{sa}n_a))}
\nonumber\\
n_a &=&\frac{1}{1+\exp(-\beta(\mu-V_a-z_a \epsilon_a n_a-z_{as}\epsilon_{sa}n_s))}
\end{eqnarray}

   	First, we consider the case where the shell-axial interparticle energy $\epsilon_{sa} = 0$,
so that we are left with 2 decoupled Ising problems. It is known that a lattice
gas can be regarded as a lattice of spins,  with the conversion
$ s=2n-1 $, $J=-\epsilon/4$ and the magnetic field $h=(\mu-V)/2-z\epsilon/4$.
One can find the chemical potential of condensation from the
condition for the magnetic transition $(h = 0)$, and the mean field critical temperature $T_c$ in
the Ising model: $\beta_{c} z J=1$, where $\beta_c=(k_B T_c)^{-1}$. In the following, we
take Boltzmann's constant $k_B =1$. Thus, $T_c=z \epsilon/4$.
Therefore, in the decoupled case, the shell and axial critical
temperatures are $T_{cs}=z_s \epsilon_s/4$ and $T_{ca}=z_a \epsilon_a/4$, respectively.
For simplicity, we again use the same axial and shell
intra-species interaction $ \epsilon_s=\epsilon_a=\epsilon$, and
scale the
temperatures with respect to $\epsilon$ . Considering a square shell lattice
$(z_s=4)$ and a $1D$ axial lattice $(z_a=2)$, we obtain $T_{cs} =1$ and
$T_{ca}=0.5$.

  	Let us consider the effect of turning on the axial-shell
interaction. The mean field results are shown in $fig.6$ for $\epsilon_{sa}=1$.
The chemical potential of
condensation is found by a Maxwell (equal-area) construction. For a large difference
between the energies (per
particle) $E_a=-(V_a-z_a/2)$ and $E_s=-(V_s-z_s/2)$, the shell and axial species
behave as in the decoupled case; two distinct
transitions occur and the transition which occurs first (at lower $\mu$) corresponds to a lower free energy.
However, in the case of similar
energies, the two species exhibit a common transition.
$T_c>1$ in this case because
the
cooperative system behaves like a single species of atoms, with a larger coordination number.

 	In order to compare our
analysis at finite $T$ with that at $T=0$, we keep $V_s$ (or
$E_s$)
fixed and vary $V_a$ (or $E_a$), so that we move on a line parallel
to the $E_a$ axis. In the finite $T$ case, we watch the resulting evolution of the
axial and shell critical transitions. There arises a convenient quantity for
characterizing this dependence; this is called $\delta$, defined by:
\begin{eqnarray}
\delta =(E_s-\epsilon_{sa}z_{sa}/2)-(E_a- \epsilon_{sa}z_{as}/2)
\end{eqnarray}
 	
 	 The evolution
of these transitions with $\delta$ is shown in $Fig.7$ for three
different interaction strengths. Consider first the strong attractive case 
$(\epsilon_{sa}=1)$. For small axial energies per particle $(\delta<-4)$, the shell
condensation occurs at a lower value of $\mu$ than that associated with
full condensation. The axial and shell
critical temperatures are the same as in the decoupled case ($0.5$ and $1$). This
corresponds to the $E \rightarrow S \rightarrow F$ region in $fig3$.
When $\delta=-4$, the effect of interaction between species becomes significant and the two
transitions merge. As $|\delta|$ approaches $0$,
the common transition's critical temperature  increases to the value $1.45$ (an increase of
$45\%$ ) at $\delta=0$. When $\delta$ increases from zero to $4$, $T_{c}$
decreases symmetrically with the case $\delta<0$. This
corresponds to the $E \rightarrow F$ region of $(Fig.3)$. A similar critical
temperature dependence on the difference between site binding energies
was observed in Monte Carlo simulations of benzene condensation in Na-X zeolites \cite{machta}. 
The difference was in that case, $T_c$ dropped abruptly to zero when $\delta$ exceeded
a threshold corresponding to a decoupling of the two transitions (since neither species
in that case had an infinite connected path of its own). When $\delta>4$ ,
the system returns to the case of
two separate axial and shell transitions. As the axial-shell atractive interaction
is reduced, the range of $\delta$ values corresponding to cooperative behavior decreases, as shown in $Fig.7$. Note
that the maximum value of $T_c$ for the case $\epsilon_{sa}=0.5$ is only $15\%$
greater than that of the decoupled shell transition. When $\epsilon_{sa}$ becomes very
small (0.1 in $Fig.7$), a single transition occurs for small $|\delta|$, but the
transition critical temperature equals that of the shell phase alone. 
   	
 	We have also considered the finite $T$ case of a repulsive interaction,
$\epsilon_{sa}<0$.
Again, we study the behavior with $E_s$ constant and vary
$E_a$, so
we move on a line parallel to the $E_a$ axis in $Fig.5$. The resulting isotherms, corresponding
to several different regions in $Fig.5$, are shown in $Fig.8$. A variety of scenarios
can be seen, including those with $A$ either preceding or following $S$. The behavior as a
function of $E_s$ is a logical correlate of that shown in $Fig.5$ at $T=0$.
In contrast with the attractive case, there is no $E \rightarrow F$ region, even for
similar axial and shell energies because the shell atoms, which have a higher density,
expel the axial atoms. However,
there occurs a qualitative similarity of the $T_c$ behavior. At low $\mu$, the axial
atoms condense first. Then, at higher $\mu$, the shell is occupied while the axis is
emptied. This transition occurs at the same $T_c$ as the cooperative transition
in the attractive case. When the external pressure (i.e. $\mu$) is sufficiently high
to overcome the axial-shell
repulsive interaction, a full condensation occurs. These features are expressed
in the ($T_c,\mu$) diagram for various values of
$\epsilon_{sa}$ and $\delta =0$ ($Fig.9$).

\section{REAL GASES IN CARBON NANOTUBES}

	We have discussed so far a simple and general theoretical model for adsorption of gases
 in a nanopore. Now, we consider the model's prediction for a specific case -
various gases adsorbed in C nanotubes. In the spirit of the model, we employ a number
of simplifying assumptions. The adsorption potential we use is described in
\cite{stan}; it is a sum of Lennard-Jones (LJ) two-body interactions
between the C atoms (spread into continuous matter) and the adsorbate. The energy and distance parameters of
this pair potential are obtained from semiempirical combining rules involving
the LJ parameters of the C atoms $(\epsilon_{CC},\sigma_{CC})$ and the adsorbate
$(\epsilon_{gg},\sigma_{gg})$ \cite{steele,stee,scoles}:
\begin{eqnarray}
\epsilon_{gC}&=& \sqrt(\epsilon_{gg}\epsilon_{CC})
\nonumber\\
\sigma_{gC}&=&(\sigma_{gg}+\sigma_{CC})/2
\end{eqnarray}
                                                                              
 The potential in the nanotube interior at distance $r$ from the
axis of the cylinder is \cite{george}:
\begin{equation}
V(r,R) = 3\pi \theta \epsilon_{gC}\sigma_{gC}^2 [\frac{21}{32}(\frac{\sigma_{gC}}{R})^{10} M_{11}(\frac{r}{R})-(\frac{\sigma_{gC}}{R})^4 M_5(\frac{r}{R})]
\end{equation}
where $R$ is the nanotube
radius, $\theta = 0.32 A^{-2}$ is the surface density of graphene $C$ atoms and
\begin{equation}
M_n(x)=\int_{0}^{\pi} \frac{d\phi}{(1+ x^2 - 2x\cos(\phi))^{n/2}}
\end{equation}

	The adsorption model is simple: the adatoms condense in a close-packed
configuration, in both the shell and axial phases. We are excluding the case of very
large $R$, which would result in the possibility of several concentric shells.
As discussed in sections $2$ and $3$, our model has $4$ parameters:
the shell and axial energies, the ratio of densities $(\gamma)$ and the inter-species interaction
$(\epsilon_{sa})$. They are not completely independent.
One can readily identify the axial potential energy
as $V_a=V(0,R)$. To find the shell potential, one should examine the form of potential.
If $R$ is large, $V(r,R)$
has a minimum for a radius $R_0$ larger than the hard-core adsorbate radius
$\sigma_{gg}$; then it is logical to assume that the gas atoms will be adsorbed in the
shell phase at this distance $(R_s=R_0)$ and the shell potential is $V(R_0,R)$. If the pore radius is small
($R_0<\sigma_{gg}$),
it is convenient to
identify $R_s=\sigma_{gg}$ and the shell potential $V_s=V(\sigma_{gg},R)$.
Geometrical calculations show that this is a good approximation,
assuming that shell atoms sit near the optimal distance $r_{min}=2^{1/6}\sigma_{gg}$
from axial atoms. There is arbitrariness in these assignments, a situation which is
inherent in any lattice model. $V_a$ and $V_s$ lead easily
to the axial and shell energies per particle $E_a=-(V_a-z_a/2)$ and
$Es=-(V_s-z_s/2)$.
 	
 	The intra- and inter-species interactions are
found using Lennard-Jones parameters for the specific gas. The intra-species
interaction energy is taken as $\epsilon_{gg}$ and the inter-species energy is the
adsorbate-adsorbate interaction at $r=\sqrt[]{R_s^2+(r_{min}/2)^2}$.
\begin{equation}
\epsilon_{sa}=4\epsilon_{gg}((\frac{\sigma_{gg}}{r})^{12}-(\frac{\sigma_{gg}}{r})^6)
\end{equation}	

  	The number of shell atoms contained in a ring of radius $R_s$
is $2\pi R_s/ \sigma_{gg}$ and the corresponding number of axial atoms is 1.
Thus, an estimate of the ratio of densities is:
\begin{eqnarray}
\gamma=\frac{N_a}{N_s}=\frac{\sigma_{gg}}{2\pi R_s}
\end{eqnarray}

  	Table $1$ presents the resulting values of the various parameters for $H_2$ and
$Xe$ inside nanotubes of various radii. The sequence of transitions is
based on data in $Fig.3$.
	We note several features of these results. First,
the only predicted transition scenarios are $E\rightarrow S\rightarrow F$, $E\rightarrow F$,
$E\rightarrow A$
and no transition. 
The $E\rightarrow S$ and $E\rightarrow A\rightarrow F$ sequences are not found for $H_2$
or $Xe$. Physically, $E\rightarrow S$
corresponds to an attractive shell potential (negative $V_s$) but a repulsive
axial potential (positive $V_a$); and $E\rightarrow A\rightarrow F$ corresponds to
very attractive axial potentials and less attractive shell potentials.
These do not occur in our model of nanotubes.
We do
find $E\rightarrow A$ and $E\rightarrow S\rightarrow F$ transitions for a relatively
large
range of nanotube radii. The cooperative behavior $E\rightarrow F$ occurs for
a very small range of parameters because the gas-gas interaction strength is weak in
comparison with the nanopore attraction. However, in the case of $Xe$, which has a much bigger
cohesive energy ($\epsilon_{gg}=221 K$) than $H_2$ ($\epsilon_{gg}=37 K$), the mutual
transition is more common. The $H_2$ gas undergoes the
$E\rightarrow S\rightarrow F$
transitions for nanotubes with $R>6$ $\mathring{A}$ , whereas $Xe$
goes through these transitions only for $R>7.7$ $\mathring{A}$. This is due to the difference
between these molecules' sizes and
interaction strengths. For $R<7.3$ $\mathring{A}$, $Xe$ can accommodate
only the axial phase, whereas the $H_2$ gas would go in the axial phase for
$R<5.8$ $\mathring{A}$. For very small $R$ (3.5 $\mathring{A}$ for $Xe$,
3 $\mathring{A}$ for $H_2$), gas does not adsorb at all in nanotubes because the pore gas
potential becomes repulsive .
	
  Hartree model calculations and path integral simulations were previously performed for
adsorption of $H_2$ in $C$ nanotubes of radii $6,7$ and $8$ $\mathring{A}$ \cite{gati}. Our classical
results are in qualitative agreement with these results. The previous study
also found the $E\rightarrow S\rightarrow F$ for this range
of nanotube radii. However, their quantum calculations allowed them to investigate the
delocalization of the axial state. For $R=8$ $\mathring{A}$ the axial state's probability density
is no longer confined to the immediate vicinity of the axis, exhibiting a
maximum near $r=2$ $\mathring{A}$. This is actually not an axial phase, but rather a second 
shell phase, of small radius. In our
calculations, the axial phase is confined to the nanotube axis and such a
second shell phase is not considered.

\section{SUMMARY AND CONCLUSIONS}

  We have investigated the adsorption of gases in nanopores, employing a lattice model,
which we solved exactly at $T=0$ and approximately at finite $T$.
Various regimes of transition behavior 
were found, corresponding to
a range of interaction strengths. The sequence of
transitions as a function of $\mu$ depends on both the axial-shell interaction energy and the
difference between the axial and shell energies per particle. When this difference is large, the two
species condense independently, i.e. the two species are essentially decoupled. When this difference is
small, the behavior depends on the sign of the axial-shell interaction.
For $\epsilon_{sa}>0$ (attractive case), the axial and shell phases
undergo a common transition at a higher critical temperature. For $\epsilon_{sa}<0$,
an increase of the critical temperature occurs, corresponding to an $A\rightarrow S$
transition.  

 The most important parameter
is the radius of the nanotube or, specifically ,the ratio $R^*$ discussed in the
introduction. Even though its value does not appear explicitly in section $2$
and $3$, it determines most of the other parameters. This is discussed in
section $4$, where the $R$ dependence of the behavior is explored. Depending on the
adsorbate size and interaction
strength, we find typically that $E\rightarrow A$ occurs for small $R^*$,
the $E\rightarrow S\rightarrow F$ occurs for large $R^*$ and the coupled condensation
($E\rightarrow F$) occurs for a small range of intermediate $R^*$.

  Our approach certainly oversimplifies the real situation in nanopores. First, the
lattice gas model
constrains the atoms to artificial sites that must be identified only by a
very approximate ansatz, discussed in Section 4. For light gases, such as
$H_2$ and $He$, quantum effects (such as zero point motion) are very important, yet
they are neglected here.
Nevertheless, we think that our model yields the principal qualitative features of
the adsorption's dependence on the various interactions
present in this problem. Thus it should help us understand the evolution of adsorption
phenomena as a function of adsorbate and pore radius.

	This research has been supported in part by grants from the Petroleum Research Fund
of the American Chemical Society and the Army Research Office.

\begin{center}
\begin{table}[h]
\caption{Possible transitions for different gases and nanotube radii.
The Lennard-Jones parameters are : $\sigma_{gg}=3.05$ $\mathring{A}$, $\epsilon_{gg}=37$ $K$
 for $H_2$ and $\sigma_{gg}=4.1$ $\mathring{A}$, $\epsilon_{gg}=221$ $K$ for $Xe$.
 All interaction energies ($V_s,V_a,E_s,E_a,\epsilon_{sa}$) are expressed
in units of the gas hard-core energies $\epsilon_{gg}$ and radii in $\mathring{A}$. The last column
shows the sequence of adsorbed phases as $\mu$ increases.}
\setlength{\tabcolsep}{0.4cm}
\renewcommand{\arraystretch}{1.5}
\begin{tabular*}{16.90cm}[t]{|c|c|c|c|c|c|c|c|c|c|}\hline

   &    $R_{nt}$& $R_s$& $V_s$& $V_a$&   $\gamma$&    $\epsilon_{sa}$& $E_s$& $E_a$& $Sequence$  \\ \hline

$H_2$&   8.0&    4.76&    -17.3&    -2.70&    0.10&    0.18&    19.3&   3.70&    $E\rightarrow S\rightarrow F$  \\
	
  &	7.0&    3.75&    -18.4&    -4.50&    0.13&    0.54&    20.4&   5.05&    $E\rightarrow S\rightarrow F$   \\
	
  & 6.0&    3.06&    -14.4&    -7.35&    0.15&    0.98&    16.4&   8.35&    $E\rightarrow S\rightarrow F$   \\

  & 5.9&    3.05&    -8.92&     -8.92&    0.16&    0.99&    10.9&    9.92&   $E\rightarrow F$ \\
  	
  &	5.8&    3.05&    +1.35&     -9.50&    0.16&    0.99&    0.64&    10.5&   $E\rightarrow A$ \\
  	
  &	5.5&    3.05&    +94.0&     -11.7&    0.16&    0.99&    -92.0&   12.7&   $E\rightarrow A$  \\

   &	3.0&    -&    -&     +1.74&    -&    -&    -&    -0.74&   $E$ \\
 	
$Xe$&    8.0&    4.20&    -10.3&    -2.70&    0.15&    0.95&    12.3&   3.70&    $E\rightarrow S\rightarrow F$  \\ \hline

   &	7.7&    4.10&    -9.97&     -3.14&    0.16&    0.99&    12.0&    4.14&   $E\rightarrow S \rightarrow F$ \\

    &	7.6&    4.10&    -8.91&     -3.30&    0.16&    0.99&    10.91&    4.30&   $E\rightarrow F$ \\

   & 	7.5&    4.10&    -7.00&     -3.50&    0.16&    0.99&    9.00&    4.50&    $E\rightarrow F$ \\

   & 	7.4&    4.10&    -4.02&     -3.67&    0.16&    0.99&    6.02&    4.67&    $E\rightarrow F$ \\
    	
   &  	7.3&    4.10&    +0.18&     -3.80&    0.16&    0.99&    1.80&    4.80&    $E\rightarrow A$  \\
     	
   &  	7.0&    4.10&    +39.0&    -4.50&    0.16&    0.99&    -37.0&  5.50&    $E\rightarrow A$ \\
     	
    & 	4.0&     -&        -&       -24.1&     -&       -&       -&     25.1&   $E\rightarrow A$  \\
     	
   &    3.5&     -&        -&       -0.50&      -&       -&       -&     -1.50&   $E$ \\ \hline

\end{tabular*}
\end{table}
\end{center}

\newpage

1. Schematic transverse section of a nanotube, showing occupied and unoccupied axial and shell sites.

2. $T=0$ phase diagram in the case of an attractive axial-shell
interaction. $\mu$ is the chemical potential and $E_s$ is defined in Eq.5. Both of
these energies are
scaled to the intra-species interaction $\epsilon$. The dashed line is the
chemical potential of bulk condensation. We distinguish two cases: (a)when the
$S \leftrightarrow F$ transition is present $(-E_a+\epsilon_{sa}/\gamma>-3)$
and (b)when the $S \leftrightarrow F$ transition is absent.

3. $T=0$ phase diagram showing the sequence of transitions
as a function of shell and axial energies, in the case of an attractive
axial-shell interaction. Arrows indicate direction associated with increasing $\mu$.

4. $T=0$ phase diagram in the case of a repulsive axial-shell interaction:
a) as a function of $E_s$ and $\mu$, with $E_a$ fixed and b) as
a function of $E_a$ and $\mu$, with $E_s$ fixed

5. $T=0$ phase diagram of possible transitions as a function of
interactions, in the case of a repulsive axial-shell interaction.

6. Adsorption isotherms in the attractive case: (a) and (c) two
transitions $T_{ca}=1$, $T_{ca}=0.5$ (as in the decoupled case) occur for
a large difference between axial and shell energies; the phase which is first
occupied corresponds to a lower energy: axial for (a), shell for (c);  (b) a
cooperative transition at a higher $T_c$ occurs when the axial and shell
energies are similar.

7. Dependence of the axial and shell critical temperatures on the
difference between axial and shell energies. The two transitions for
different $\mu$ occurring at large $\delta$ merge into one common transition
when $|\delta|<4\epsilon_{sa}$. The width of the cooperative behavior regime is
proportional to $\epsilon_{sa}$.

8. Isotherms at finite T in the case of repulsive axial-shell
interaction: (a) and (c) two transitions at the decoupled critical temperatures
occur for a large difference between the axial and shell energies. The
phase which occurs at lower $\mu$ corresponds to a lower energy per site. In (b) three different
transitions occur when the axial and shell
energies are similar.

9. Transition curves in $\mu-T$ plane at $\delta=0$ for various values of
$\epsilon_{sa}$, in the case of a repulsive axial-shell interaction. The axial
sites are filled first; then, when the shell gets filled, axial atoms are expelled
and finally, as $\mu$ increases, the full phase occurs. As in the
attractive case, the critical temperature is enhanced by the coupling.

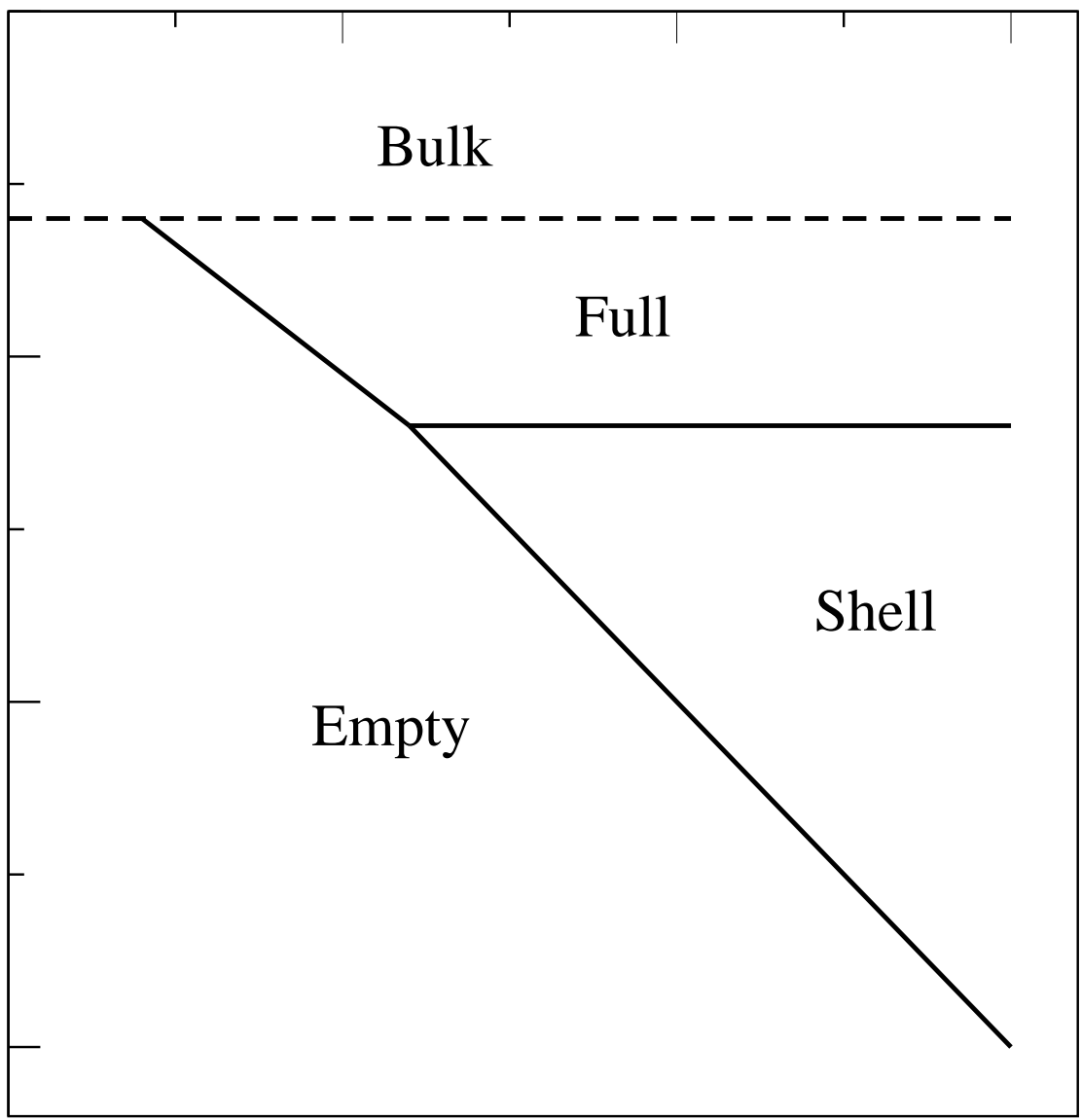
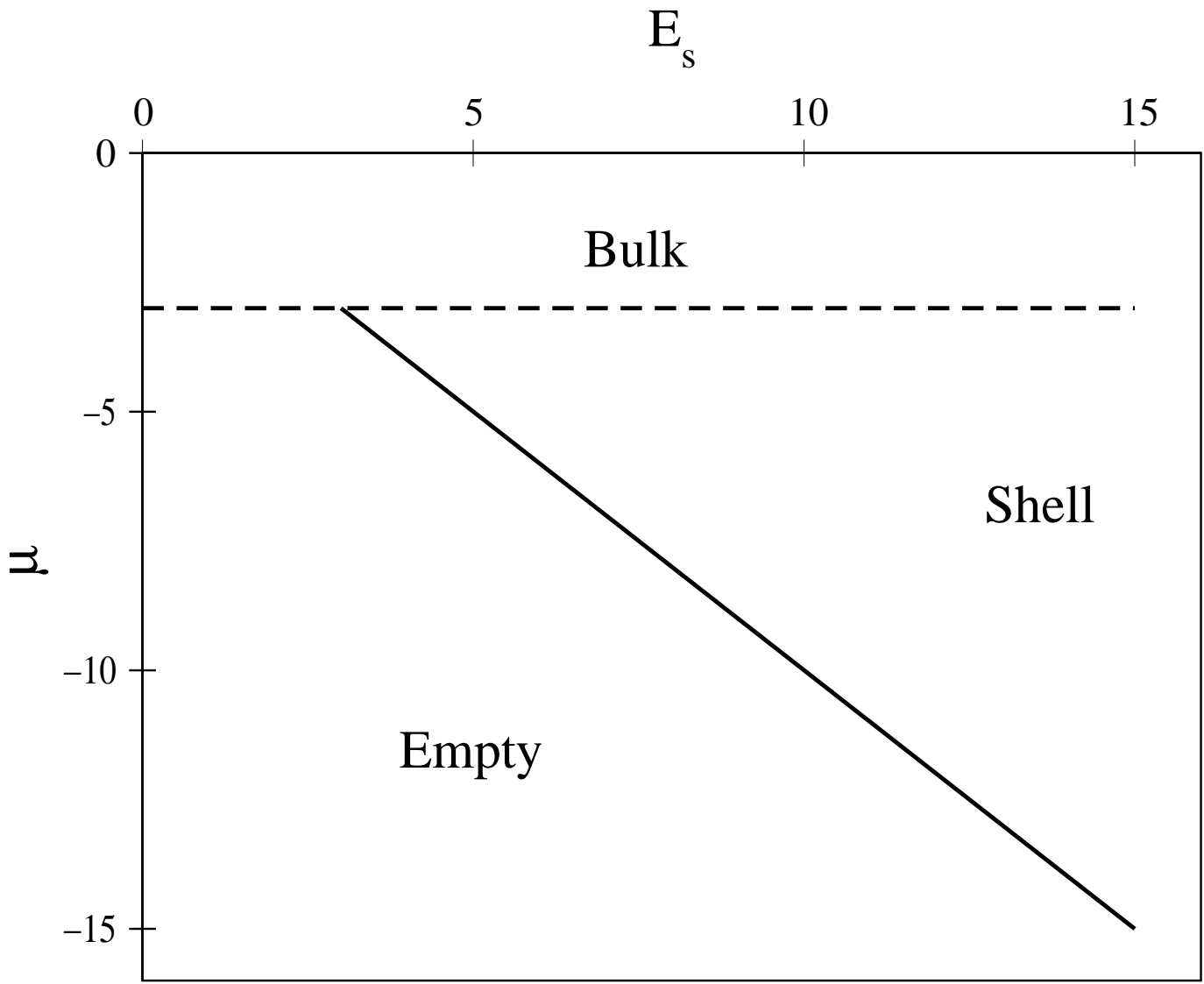
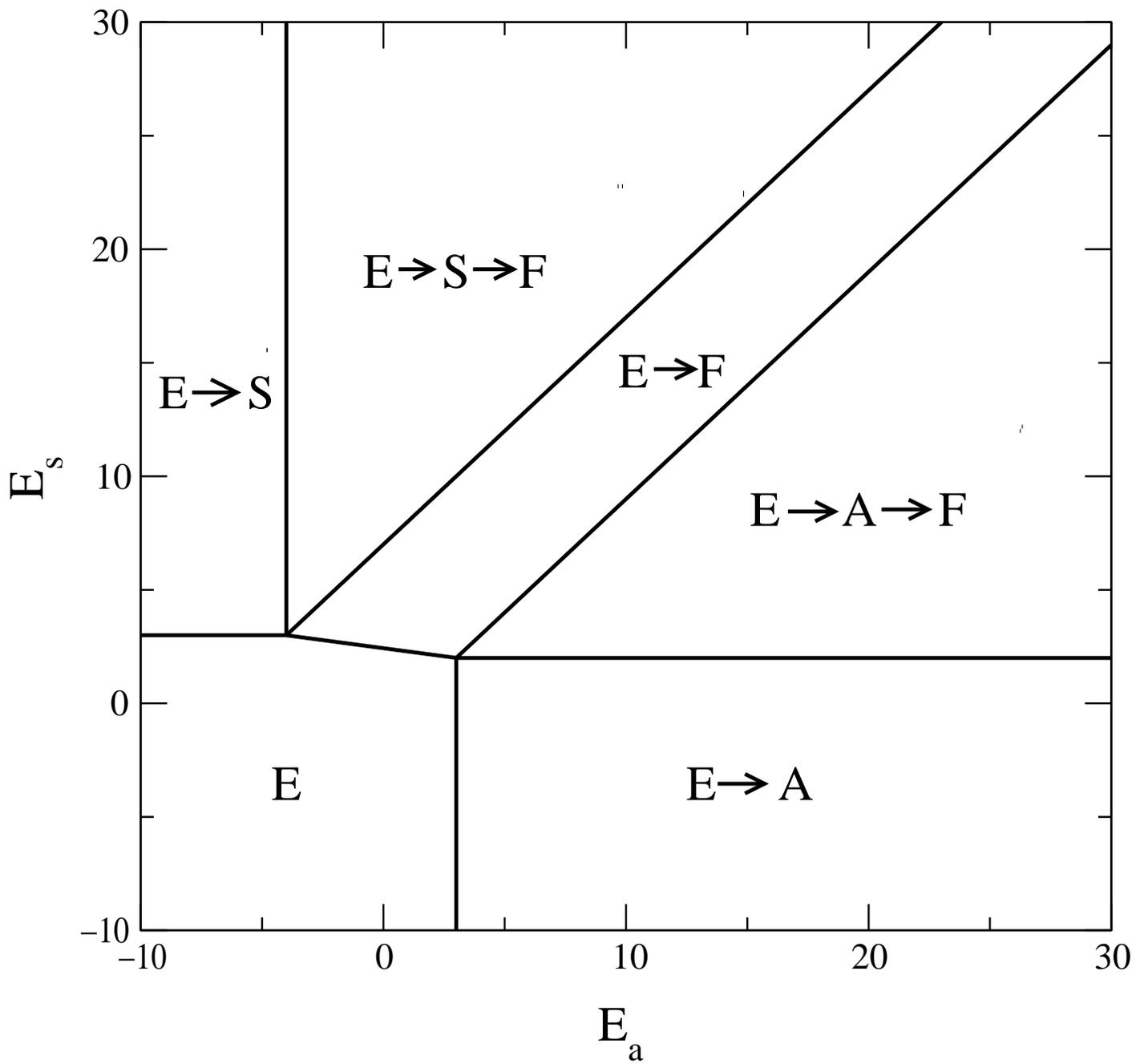
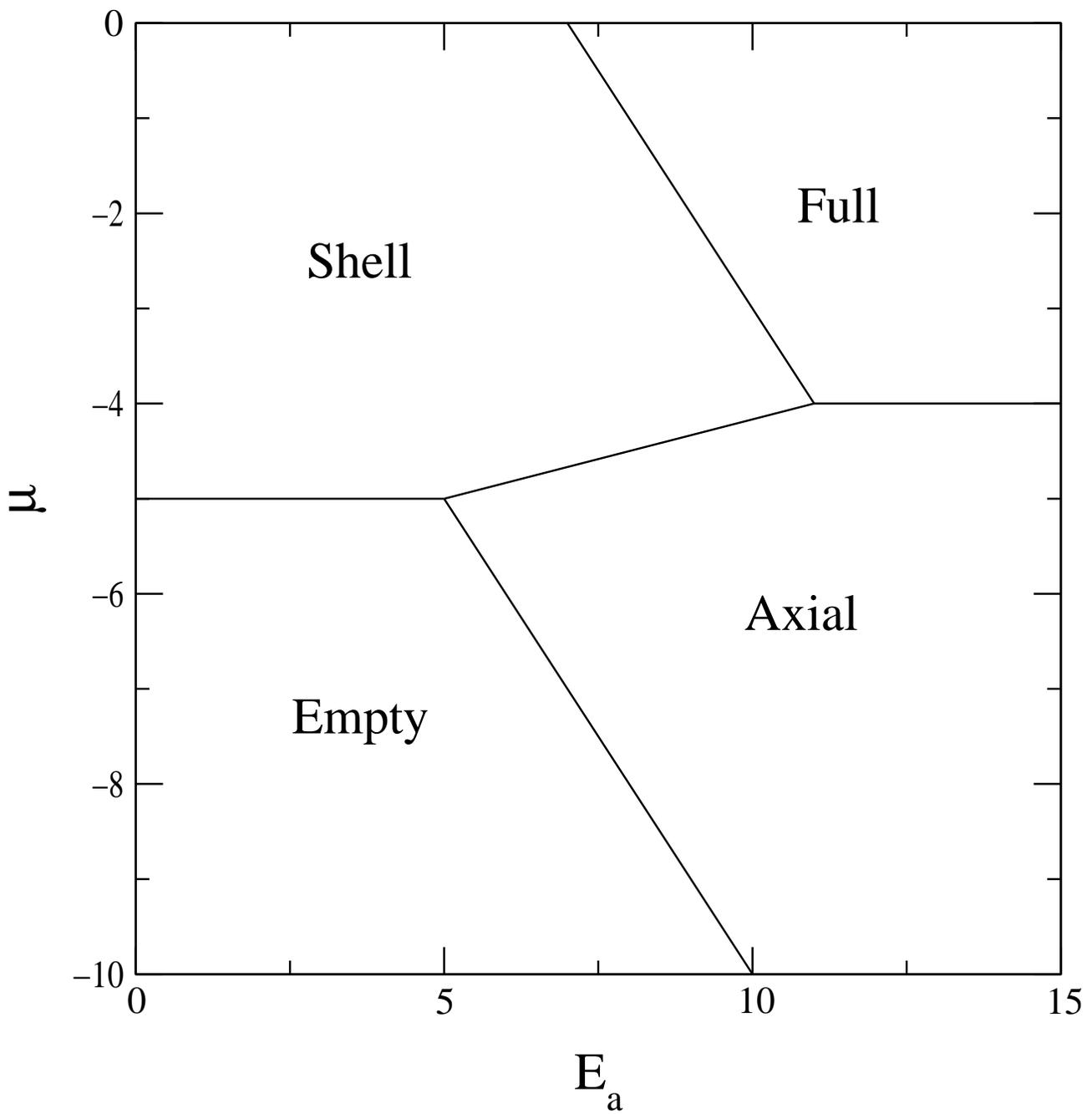
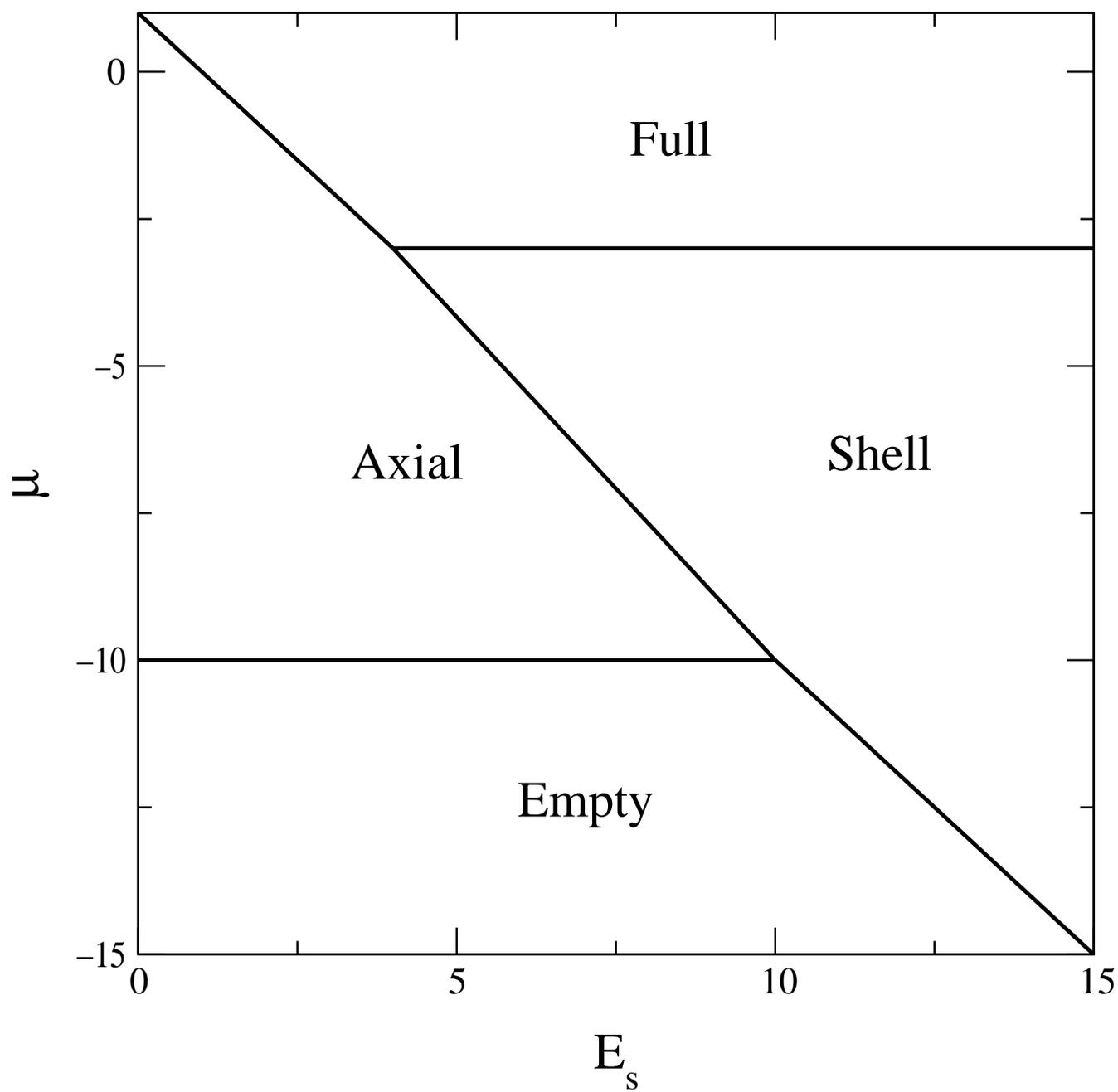
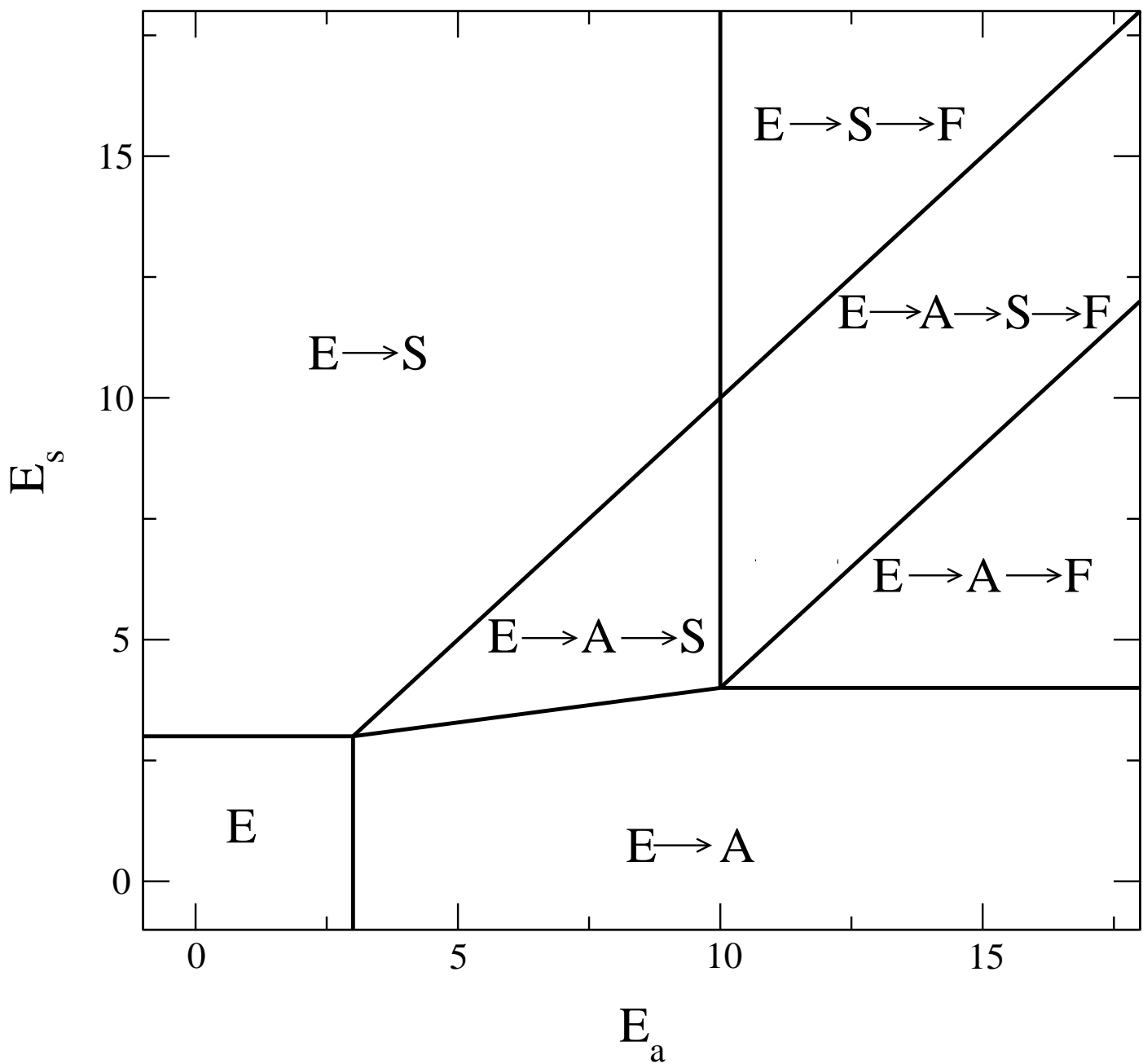
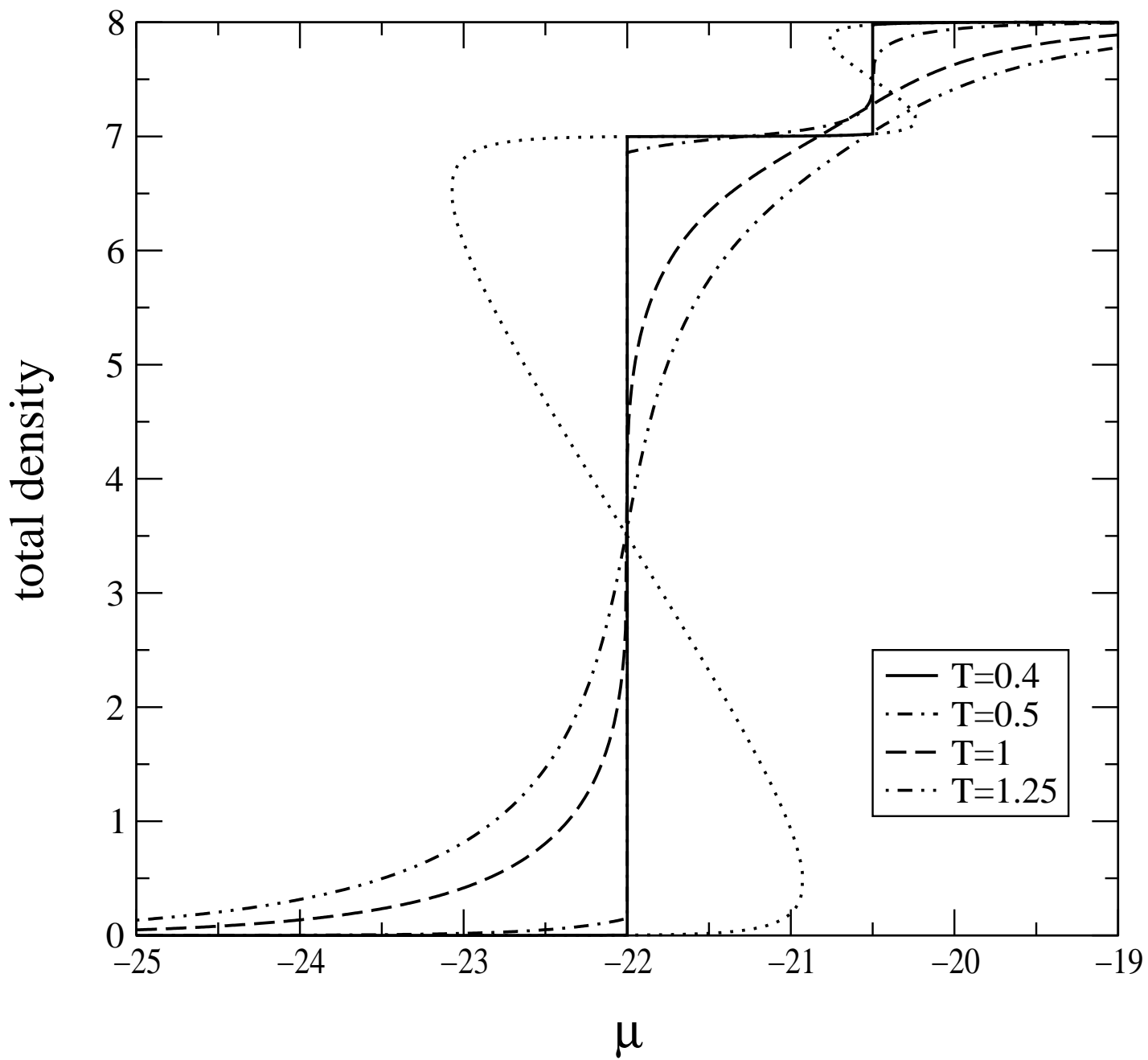
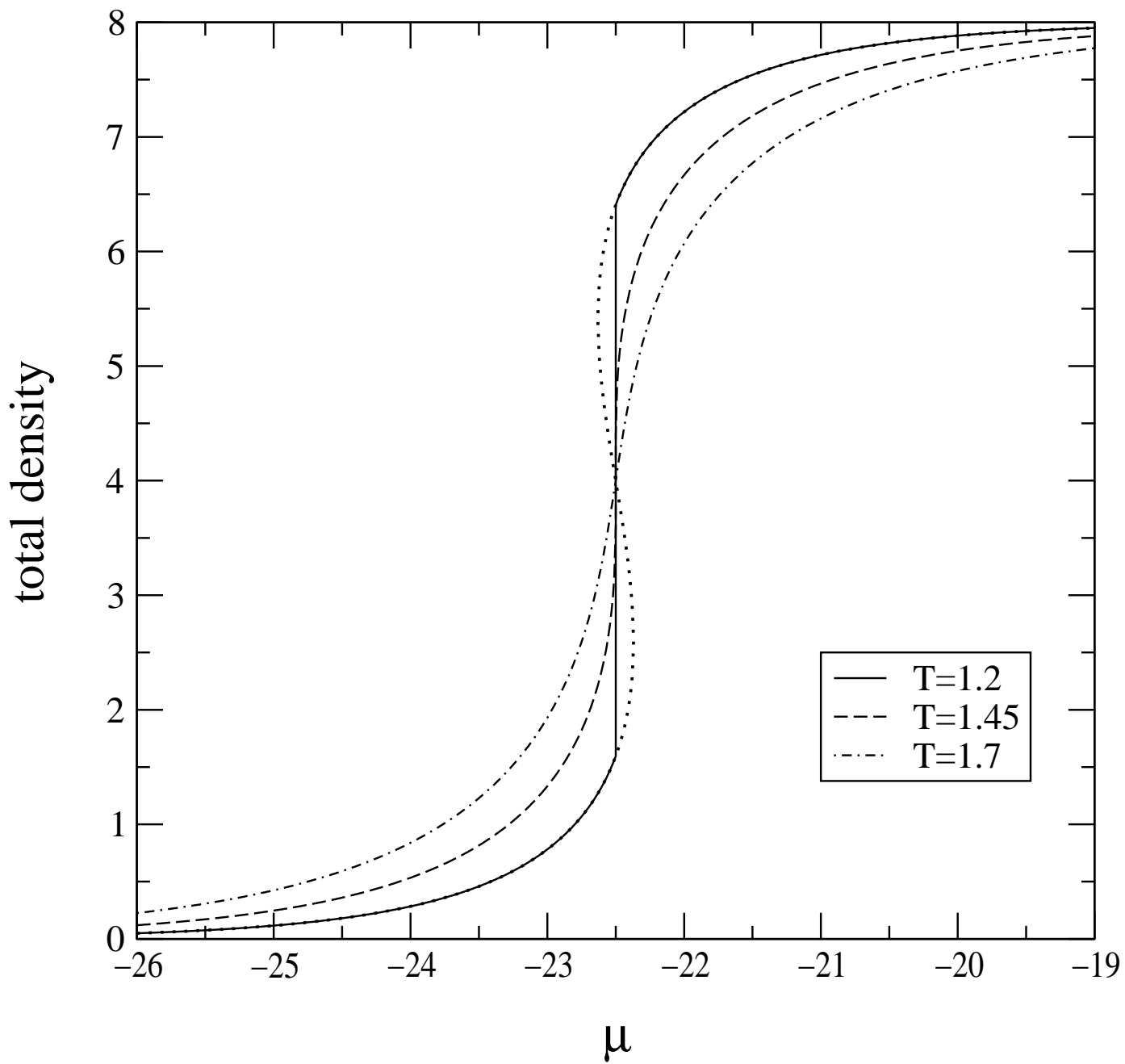
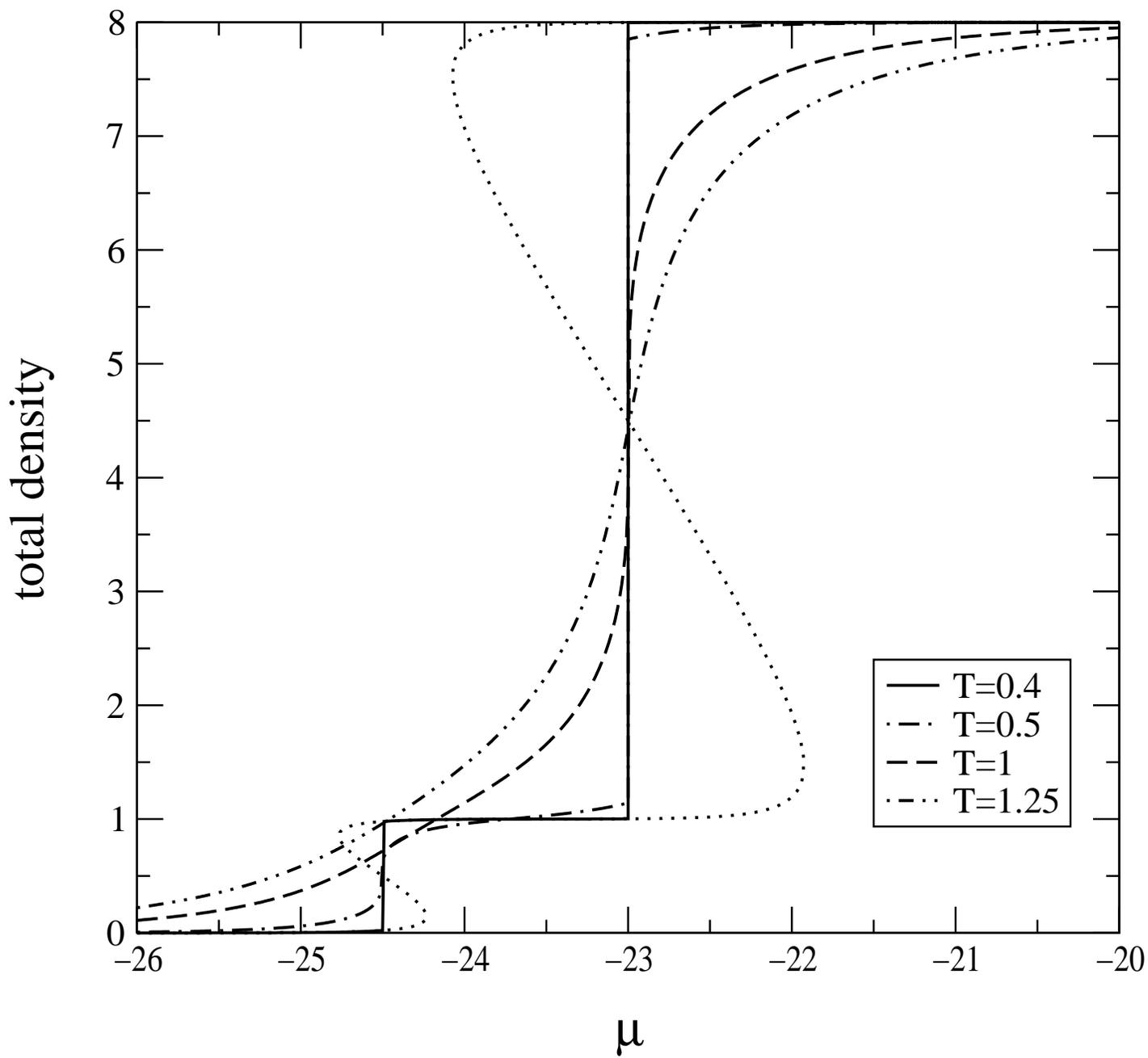
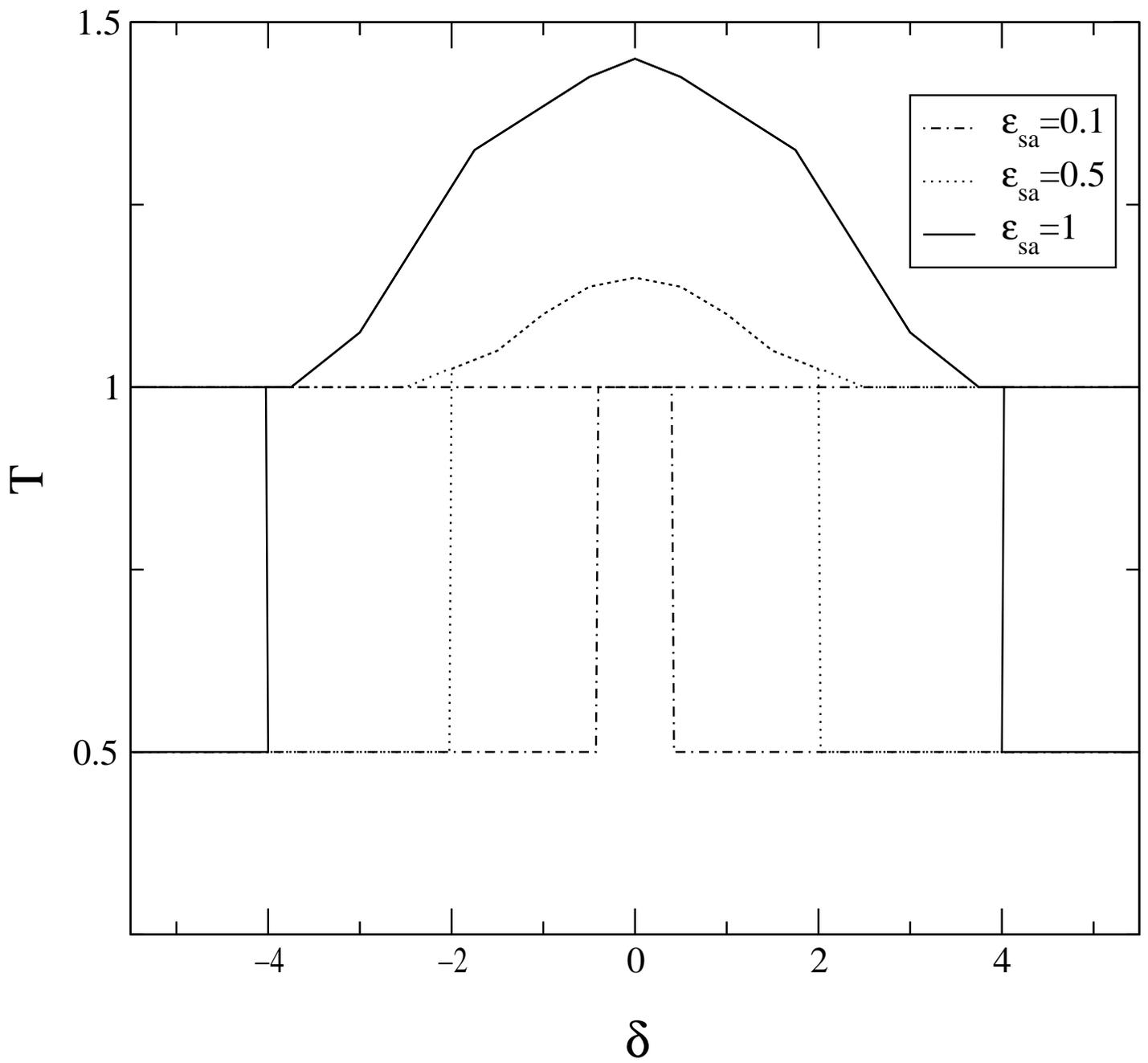
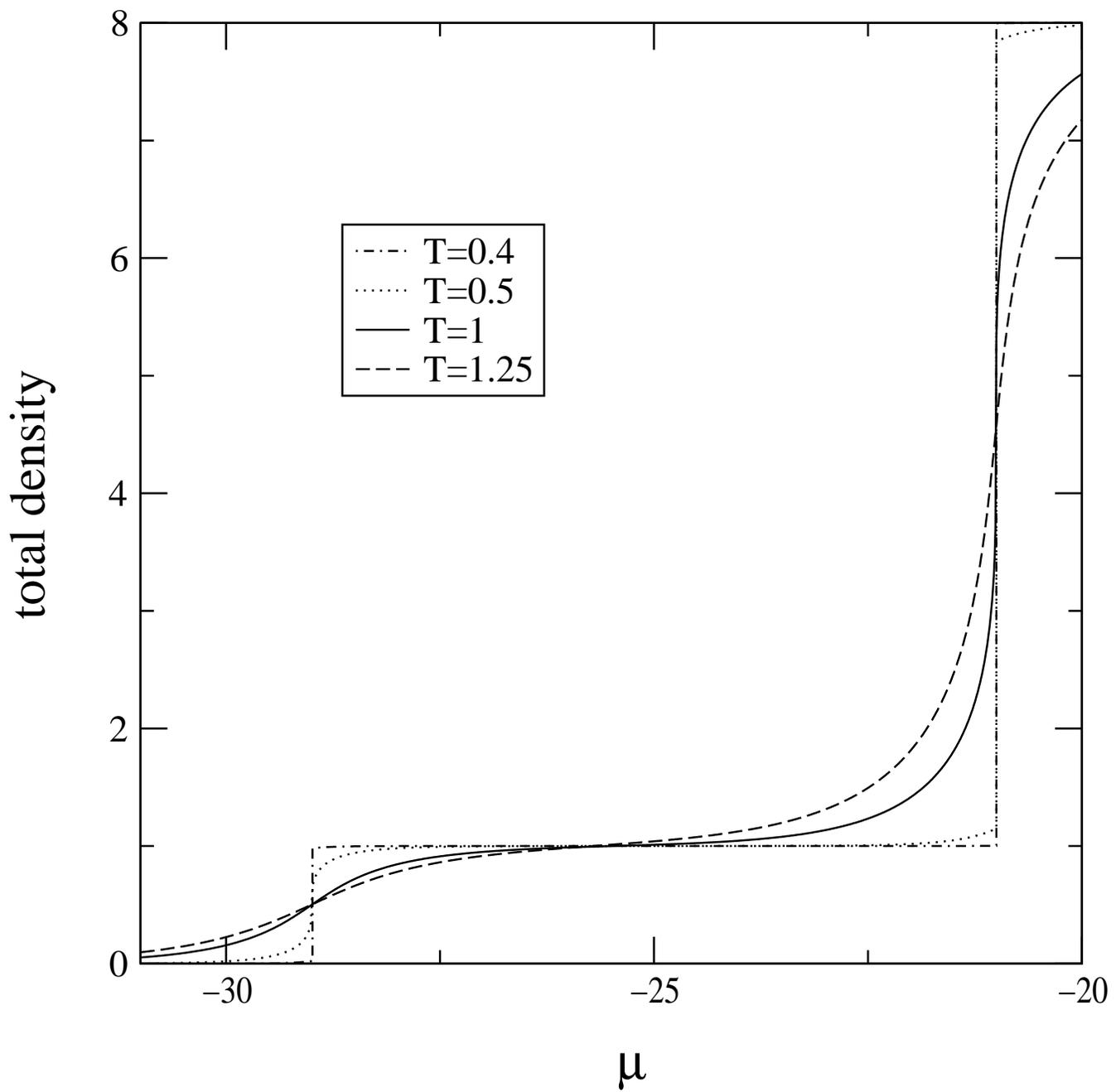
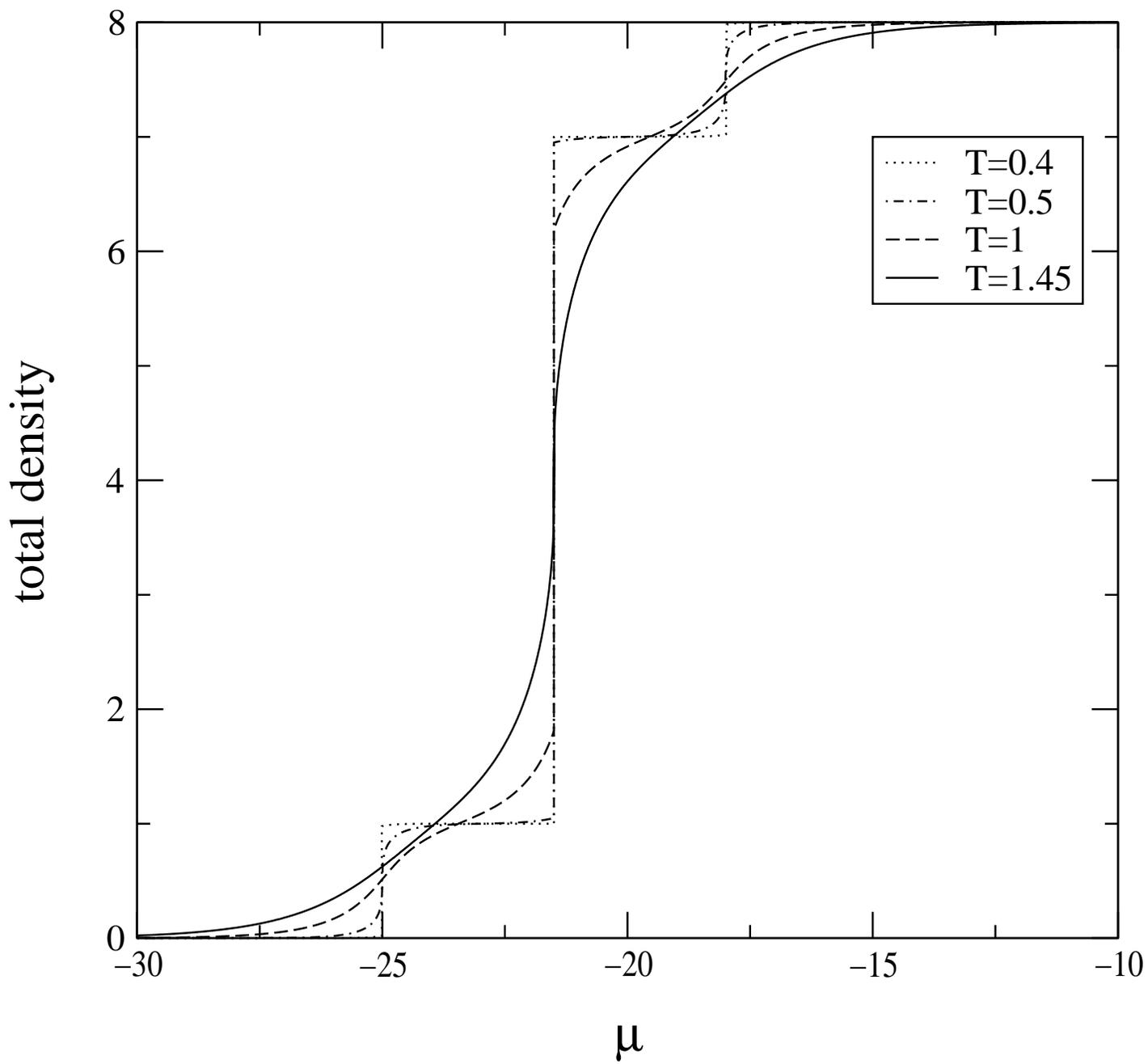
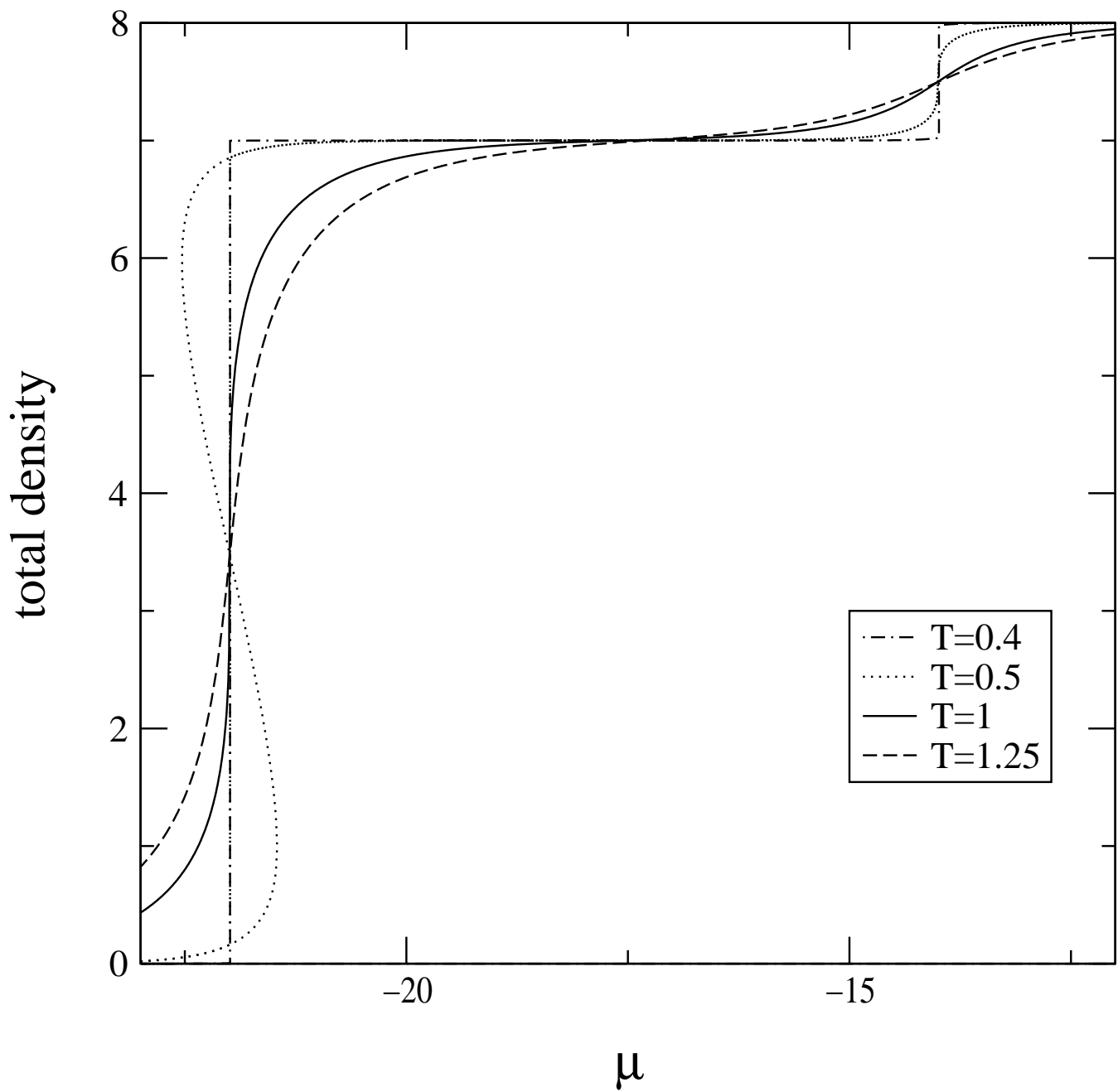
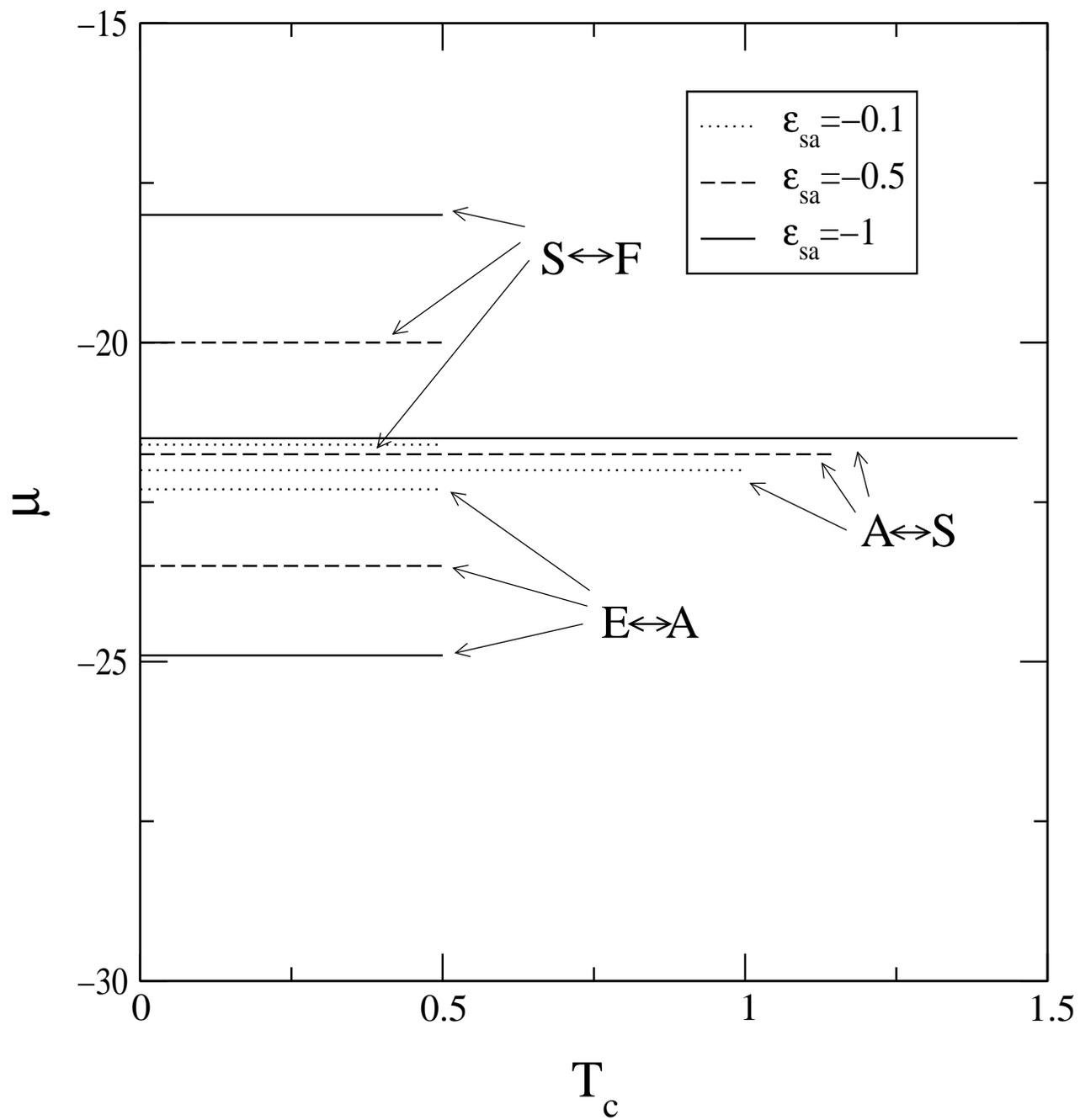
\end{document}